\documentclass[12pt,onecolumn]{aastex6}

\usepackage{epsfig}

\usepackage[caption = false]{subfig}

\usepackage{amsmath}
\usepackage{graphicx}
\usepackage{float}
\usepackage{amsfonts}
\usepackage{amssymb}
\usepackage{makeidx}
\usepackage{mathrsfs}
\usepackage{flushend}

\newcommand{\sigmaT}{\sigma_{_{\rm T}}}
\newcommand{\lapprox}{\lower.4ex\hbox{$\;\buildrel
<\over{\scriptstyle\sim}\;$}}

\newcommand{\gapprox}{\lower.4ex\hbox{$\;\buildrel
>\over{\scriptstyle\sim}\;$}}

\begin{document}

\title{TIME-DEPENDENT ELECTRON ACCELERATION IN PULSAR WIND TERMINATION SHOCKS: APPLICATION TO THE 2007 SEPTEMBER CRAB NEBULA GAMMA-RAY FLARE}

\author{John J. Kroon}

\affil{National Research Council, resident at the Naval Research Laboratory; Washington, DC 20375, USA; john.kroon.ctr@nrl.navy.mil; jkroon@gmu.edu}

\author{Peter A. Becker}

\accepted{for publication in ApJ}

\affil{Department of Physics and Astronomy, George Mason University, Fairfax, VA 22030-4444, USA;pbecker@gmu.edu}

\author{Justin D. Finke}

\affil{U.S.\ Naval Research Laboratory, Code 7653, 4555 Overlook Ave.\ SW,
        Washington, DC,
        20375-5352; 
        justin.finke@nrl.navy.mil\\
}

\begin{abstract}

In 2007 September, the Crab Nebula exhibited a bright $\gamma$-ray flare in the GeV energy range that was detected by {\em AGILE}. The observed emission at $\ga 160$\ MeV indicates that the radiating electrons had energies above the classical synchrotron radiation-reaction limit, thus presenting a serious challenge to classical models for electron acceleration in astrophysical environments. In this paper, we apply our recently developed time-dependent self-similar analytical model describing electrostatic acceleration in the explosive reconnection region around the pulsar wind termination shock to the 2007 September flare. This event was unique in that it displayed both long-duration ``wave'' and short-duration ``sub-flare'' features. The unusual temporal variation makes this flare an especially interesting test for our model. We demonstrate that our model can reproduce the time-dependent $\gamma$-ray spectrum for this event, as well as the associated $\gamma$-ray light curve, obtained by integrating the spectrum for photon energies $\ge 100\,$MeV. This establishes that our time-dependent electrostatic acceleration model can explain both wave and sub-flare transients, which lends further support to the theoretical framework we have developed. We also further examine the validity of the self-similar electric and magnetic field evolution implied by our model. We conclude that strong electrostatic acceleration driven by shock-induced magnetic reconnection is able to power the Crab Nebula $\gamma$-ray flares by energizing the electrons on sub-Larmor timescales.\\

\end{abstract}

\section{INTRODUCTION}
\label{intro}

The Crab Nebula has been observed extensively since its progenitor star went supernova in 1054 CE, and in recent decades its high-energy emission has displayed a rich variety of behaviors which have been documented by a series of space-based $\gamma$-ray and X-ray observatories. It is a relatively nearby and young pulsar wind nebula that, until recently, was thought to produce stable synchrotron emission from electrons and positrons (hereafter, electrons) in the remnant's magnetosphere, and thus has been used as a calibration source \citep[for a review see][]{buehler14}. The nebula is powered by the rapidly rotating pulsar which is gradually slowing down. A portion of the resulting spin-down power, $\approx 5 \times 10^{38}\,\rm erg\,s^{-1}$ is deposited into the outward-flowing electron wind. A standing termination shock forms at the radius at which ram pressure balances the gas pressure of the nebula, corresponding to a distance $r_t \sim 10^{17}\,$cm from the pulsar \citep{rees74,montani14}.\\

The relativistic electrons in the upstream portion of the wind interact with the ambient magnetic fields and produce synchrotron radiation as they advect outwards. Particles accelerated at the termination shock can explain the broad energy distribution of the observed quiescent emission \citep{gaensler06}. The accelerated particles diffuse and advect downstream from the shock, producing the synchrotron emission observed in the outer region of the nebula. Thus, it is not surprising that the apparent angular size of the Crab Nebula increases with decreasing energy \citep{abdo11}. The oldest electrons continue to advect far downstream of the shock; the emission characterizes the radio synchrotron nebula out to a radius of $\sim 10^{18}\,$cm.\\

The 2007 September $\gamma$-ray flare event was observed by {\em AGILE} and analyzed by \citet{striani13}. They define a classification scheme for episodes of enhanced $\gamma$-ray emission in which ``sub-flares'' are very short-duration events lasting a day or less, and ``waves'' are outbursts that have longer durations and peak brightnesses about half that of the sub-flares. The light curve plotted in Figure~1 from \citet{striani13} displays a clearly-defined and isolated wave event, designated W1, which takes place roughly between 2007 September 25-30 (MJD 54368-54373) and is followed by a second wave, W2. However, in the case of W2, after the wave reaches its peak, it is interrupted by a series of bright and short-lived sub-flares which are designated as F1, F2 and F3. The authors present a time-averaged spectrum for the W1 event over the dates in which it is defined, and a time-averaged spectrum for the W2 event up to its peak only. They also provide a spectrum for the peak of the F2 sub-flare event, integrated over 12-hour time bins. Based on the time constraints associated with the various spectral features that they define, we find it convenient to focus on analysis of the F2 sub-flare and the W1 wave event, since these have the highest quality spectral data.\\

\citet[][hereafter Paper~1]{kroon18} developed a time-dependent self-similar model describing the acceleration and energy losses experienced by a population of magnetically-confined relativistic electrons in the vicinity of the pulsar-wind termination shock. The observed synchrotron radiation produced by these electrons is determined by the local magnetic field strength, combined with the energy distribution of the radiating electrons. They assumed a Gaussian initial energy distribution for the incident electrons, which is a reasonable approximation for the conditions in the wind plasma just upstream from the shock \citep[][Paper~1]{cerutti12}. By combining this initial condition with a rigorous transport equation, they derived an exact analytical expression for the time-evolving energy distribution of the electron population, subject to the acceleration and losses encountered as they pass through the region of explosive magnetic reconnection surrounding the termination shock. The associated synchrotron spectrum can be computed exactly, and the model can be tested via comparisons with the $\gamma$-ray data. The synchrotron spectra computed using the model are self-consistent since synchrotron energy losses are included in the transport equation. In Paper~1 we used the time-dependent model to analyze the 2011 April Crab Nebula $\gamma$-ray event observed using the LAT detector onboard {\em Fermi}. Furthermore, an earlier, time-independent version of the model was used to analyze all of the available $\gamma$-ray spectra produced during flares observed between 2007-2013 \citep{kroon16}.\\

The paper is organized as follows. In Section~\ref{formalism} we present an abbreviated review of the time-dependent model developed in Paper~1. In Section~\ref{application} we apply the model to the analysis of the 2007 September event and discuss the resulting physical parameters. In Section~\ref{conclusion} we provide our conclusions and discuss their astrophysical significance.

\newpage

\section{Particle Transport Formalism}
\label{formalism}

In Paper~1 we modeled the 2011 April $\gamma$-ray flare using a time-dependent particle transport formalism, based on an analytical solution to a first-order transport equation, which includes terms describing electrostatic acceleration, synchrotron losses, and particle escape. The theoretical spectra computed at various time intervals compare quite favorably with the {\em Fermi} spectral data for the 2011 April event. Furthermore, the theoretical light curve for the event also agrees with the observations \citep{buehler12}. Here, we will apply the time-dependent model developed in Paper~1 to the analysis and interpretation of the 2007 September flare event observed by {\em AGILE}. The primary goal of this paper is to determine whether the same model can successfully account for the long-duration ``wave'' events observed during that transient \citep{striani13}.\\

Our model is similar to the ``blob'' picture proposed by \citet{zrake16}. In this scenario, the GeV synchrotron emission observed during the flares is produced by magnetically-confined populations of relativistic electrons that are processed via particle acceleration, losses, and escape occurring in the active reconnection region surrounding the pulsar-wind termination shock. Individual sub-flares and waves are attributed to separate blobs of plasma that independently encounter the shock. Since little is known about the precise nature of the upstream pulsar wind, we will assume that the energy distribution of the electrons in the blob as it impinges on the shock can be approximated using a broad Gaussian function whose mean ($\mu$) and standard deviation ($\sigma$) are treated as free parameters \citep[][Paper~1]{cerutti12}. Our time-dependent model allows us to follow the detailed evolution of the electron energy distribution as the particles experience strong electrostatic acceleration and synchrotron losses in the vicinity of the shock. In the remainder of Section~\ref{formalism} we provide a brief review of our model, and refer interested readers to Paper~1 for details.

\subsection{Particle Transport Equation}
\label{TransEqn}

We can model the evolution of the energy distribution of the blob electrons during a wave or sub-flare event using a spatially-averaged, time-dependent transport equation that includes terms describing electrostatic acceleration, synchrotron losses, and particle escape,
\begin{equation}
\frac{\partial f}{\partial t} = \frac{-1}{p^2}\frac{\partial}
{\partial p}\left\{p^2\left[A(t)m_e c -S(t) \frac{p^2}{m_e c} \right]f \right\}
- \frac{f}{t_{\rm esc}(p,t)}
\label{eq1}
\ ,
\end{equation}
where $p$ is the electron momentum, $t$ is time, $m_e$ is the electron mass, $c$ is the speed of light, and $f$ denotes the electron momentum distribution function. The mean time for particles to escape from the blob is represented by $t_{\rm esc}$, which can in principle depend on both $p$ and $t$. This is further discussed in Section~\ref{partesc}. The (time-dependent) total number of electrons contained in the blob is computed from the distribution function $f$ using
\begin{equation}
N_{\rm tot}(t) = \int_0^\infty 4 \pi \, p^2\, f(p,t) \, dp \ .
\label{eq2}
\end{equation}
Note that particle escape tends to decrease the number of particles in the blob over time since we assume that there is no continual particle injection once the population of electrons is initialized.

\subsection{Particle Escape Mechanisms}
\label{partesc}

In the one-zone model considered here, the electron population represents a spatial average over the active region surrounding the shock, where explosive reconnection and particle acceleration occurs. Following \citet{steinschlick}, we characterize each region according to the dominant mechanism governing the escape of particles from the active region. On the upstream side, the process known as ``shock-regulated escape'' (SRE) is thought to dominate particle escape (see Paper~1). In this scenario, higher-energy particles (with large Larmor radii) have a higher probability of being recycled back to the upstream side of the shock through the shock. Conversely, lower-energy particles will almost with small Larmor radii are swept downstream with the bulk flow. The mean escape timescale for the SRE process is given by \citep{kroon16}
\begin{equation}
t_{\rm esc} = t_{\rm SRE} = \dfrac{p}{C(t) m_e c}
\ ,
\label{tesc1}
\end{equation}
where $p$ is the particle momentum and $C(t)$ is a time-dependent parameter that controls the strength of the SRE process based on physical conditions such as the shock obliquity.\\

Once the plasma has moved through the termination shock, the dominant mechanism governing the escape of particles from the active reconnection zone surrounding the shock switches to an advection-dominated mode. In this region, even the particles with large Larmor radii are no longer recycled through the shock for additional acceleration, and instead they are swept away by the tangled magnetic field into the outer region of the nebula. Hence, on the downstream side of the shock, the escape timescale should be interpreted as the mean time for electrons to be advected away from the active zone around the termination shock. In this case, the escape timescale is independent of both energy and time, and we can write
\begin{equation}
t_{\rm esc} = t_{\rm ad} = \dfrac{R_{\rm b}}{v_{ds}}
\label{tesc2}
\ ,
\end{equation}
where $R_{\rm b}$ is the blob radius and $v_{ds}=c/3$ denotes the downstream flow velocity \citep{achterberg01}. Our model accounts for the variation in the dominant particle escape mechanism by incorporating two different formulations for the escape timescale, with Equation~(\ref{tesc1}) utilized in the upstream region and Equation~(\ref{tesc2}) utilized in the downstream region. This allows us to more accurately model the evolution of the electron distribution during the $\gamma$-ray flares within the confines of a spatially-averaged model. The upstream and downstream regions are temporally associated with the rising and decaying phases of the sub-flare (or wave), respectively.

\subsection{Profile Function and Self-Similarity}

In our self-similar time-dependent model (Paper~1), the electrostatic acceleration and synchrotron loss processes are parameterized by the functions $A(t)$ and $S(t)$ in Equation~(\ref{eq1}), which are related to the time-dependent electric and magnetic fields, $E(t)$ and  $B(t)$, respectively, via
\begin{equation}
A(t) = \frac{q E(t)}{m_e c}=A_* h(t) \ , \qquad S(t) = \frac{\sigmaT B^2(t)}{6\pi m_e c}=S_* h(t) \ ,
\label{eq3}
\end{equation}
where $q$ is the magnitude of the electron charge, the subscript ``$*$'' denotes the initial value of a quantity measured at the beginning of the $\gamma$-ray flare (at time $t=t_*$), and $h(t)$ represents the self-similar ``profile function,'' which parameterizes the variation of the electric and magnetic fields due to reconnection at the termination shock. We assume that the profile function, $h(t)$, varies according to \citep{buehler12}
\begin{equation}
h(t) =
\begin{cases}
e^{\alpha \, t/t_{\rm pk}}, & {t \le t_{\rm pk}} \ , \\
e^{\alpha} e^{-\theta (\frac{t}{t_{\rm pk}}-1)}, & {t \ge t_{\rm pk}} \ ,
\end{cases}
\label{eq4}
\end{equation}
where $\alpha$ and $\theta$ are the rising and decaying time constants, respectively, and $t$ and $t_{\rm pk}$ are measured with respect to the origin of the event at time $t_*$. This functional form for $h(t)$ was successfully used to fit the spectral data for the 2011 April flare from the Crab Nebula, and it will also allow us to reproduce the 2007 September event of interest here. It follows that the minimum value of the profile function, at the beginning of the event, is $h(t_*)=1$, and the maximum value, which occurs at the peak of the sub-flare or wave, is $h(t_{\rm pk})=e^\alpha$. Based on Equations~(\ref{eq3}), we can express the initial values of the functions $A(t)$ and $S(t)$ in terms of the initial electric and magnetic fields, $E_*$ and $B_*$, respectively, obtaining
\begin{equation}
A_* = \frac{q E_*}{m_e c} \ , \qquad S_* = \frac{\sigmaT B^2_*}{6\pi m_e c} \ .
\label{eq5}
\end{equation}

Combining Equations~(\ref{eq3}) and (\ref{eq5}), one finds that in the self-similar model, the time variations of the electric and magnetic fields are given by
\begin{equation}
E(t) = E_* \, h(t) \ , \qquad B(t) = B_* \, \sqrt{h(t)} \ ,
\label{eq6}
\end{equation}
which implies that
\begin{equation}
E(t) \propto B^2(t)
\label{eqEB2}
\ .
\end{equation}
Some useful insight on the validity of this self-similar field variation is provided by the work of \citet{lyutikov17,lyutikov18} and \citep{cerutti14}. These authors perform particle-in-cell (PIC) simulations to study the effect of explosive reconnection in plasmas with $\sigma \gg 1$. We note that Figure 7 from \citet{lyutikov17} indicates that the reconnection velocity $v_{\rm rec}$ increases approximately in proportion to the square-root of time during the early phase of the reconnection event, and therefore we can write\\ 
\begin{equation}
v_{\rm rec} \propto t^{1/2}
\label{eqvrec}
\ .
\end{equation}
Furthermore, it follows from Equation~(4.2) in \citet{lyutikov17} that
\begin{equation}
E \propto v_{\rm rec} \, B
\label{eqEvB}
\ .
\end{equation}
Figure 7 from \citet{lyutikov17} also indicates that
\begin{equation}
E \propto t
\label{eqEt}
\ ,
\end{equation}
early in the flare's evolution. Combining Equations (\ref{eqvrec}), (\ref{eqEvB}), and (\ref{eqEt}), we find that $B \propto t^{1/2}$, and consequently $E \propto B^2$, in agreement with Equation~(\ref{eqEB2}). This supports the plausibility of the self-similar relationship between $E$ and $B$ implied by our model.

\subsection{Fokker-Planck Equation}

Moving forward with analysis of the particle transport equation, it is convenient to transform from the momentum $p$ to the dimensionless parameter
\begin{equation}
x \equiv \frac{p}{m_e c} \ ,
\label{eq7}
\end{equation}
so that $x = \sqrt{\gamma^2-1}$, where $\gamma$ is the electron Lorentz factor. In the case of the ultra-relativistic electrons that power the Crab Nebula $\gamma$-ray flares, $\gamma \gg 1$, and therefore $x \to \gamma$. We can now express Equation~(\ref{eq1}) in Fokker-Planck form by writing
\begin{equation}
\frac{\partial N}{\partial y} = \frac{\partial^2}{\partial x^2}
\left(\frac{1}{2} \, \frac{d\sigma^2}{dy} \, N\right)
- \frac{\partial}{\partial x}\left(\frac{dx}{dy}
\, N\right) - \frac{1}{A_* h(y) t_{\rm esc}(x,y)} \, N \ ,
\label{eq8}
\end{equation}
where the dimensionless time
\begin{equation}
y(t) \equiv A_* \int_{t_*}^t h(t') \, dt' \ ,
\label{eq8b}
\end{equation}
and the electron number distribution
\begin{equation}
N(x,t) \equiv 4\pi (m_e c)^3 x^2f(x,t)
\label{eq9}
\ .
\end{equation}
The ``broadening'' and ``drift'' coefficients in Equation~(\ref{eq8}) are given by
\begin{equation}
\frac{1}{2} \, \frac{d\sigma^2}{dy} = 0 \ , \ \ \ \ \ 
\frac{dx}{dy} = 1-\hat S x^2
\ ,
\label{eq10}
\end{equation}
respectively. Since our model does not include momentum diffusion, the broadening coefficient vanishes in our application. Solving Equation~(\ref{eq8}) requires the specification of an initial condition and also a functional form for the escape timescale $t_{\rm esc}(x,y)$. As explained in Section~\ref{partesc}, we use Equation~(\ref{tesc1}) to compute $t_{\rm esc}$ in the upstream region and Equation~(\ref{tesc2}) in the downstream region. As discussed in Sections~\ref{secRisingPhase} and \ref{secDecayingPhase}, we also employ two different initial conditions when solving Equation~(\ref{eq8}) for the electron number distribution $N$ during the rising and decaying phases of each $\gamma$-ray sub-flare (or wave).

\subsection{Rising Phase}
\label{secRisingPhase}

During the rising phase of the sub-flare (or wave), one can obtain an exact solution for the electron distribution in the plasma blob by solving Equation~(\ref{eq8}) subject to a Gaussian initial condition and setting the escape time $t_{\rm esc} = t_{\rm SRE}$ using Equation~(\ref{tesc1}). The Gaussian initial condition represents the (incident) electron distribution in the upstream pulsar wind; that is
\begin{equation}
N(x,y)\Big|_{y=0} = \dfrac{J_0}{\sigma \sqrt{2\pi}}e^{\frac{-(x-\mu)^2}{2\sigma^2}} \ ,
\qquad \frac{1}{\sqrt{\hat S}} \ge x \ge 0 \ ,
\label{eqIC1}
\end{equation}
where $\mu$ and $\sigma$ represent the mean and standard deviation, respectively, and $J_0$ is a normalization factor. The total number of electrons initially contained in the plasma blob is given by
\begin{equation}
\mathscr{N}_0 \equiv J_0 \int_0^{1/\sqrt{\hat{S}}} \frac{1}{\sigma\sqrt{2\pi}} e^{\frac{-(x-\mu)^2}{2\sigma^2}} \, dx
= \dfrac{J_0}{2}\left[{\rm Erf}\left(\frac{\mu}{\sqrt{2}\sigma}\right) - {\rm Erf}\left(\frac{\mu-1/\sqrt{\hat{S}}}
{\sqrt{2}\sigma}\right)\right]
\label{eqIC2}
\ .
\end{equation}
The corresponding total initial energy of the electrons, $\mathscr{E}_0$, is computed by numerically evaluating the integral 
\begin{equation}
\mathscr{E}_0 = J_0 \int_0^{1/\sqrt{\hat S}}\frac{m_e c^2\sqrt{x^2+1}}{\sigma\sqrt{2\pi}} e^{\frac{-(x-\mu)^2}{2\sigma^2}} \, dx
\label{eqIC3}
\ .
\end{equation}
By combining these relations, we find for the mean initial Lorentz factor for the electrons in the plasma blob is given by
\begin{equation}
\bar\gamma_0 = \sqrt{\bar x_0^2+1} = \dfrac{\mathscr{E}_0}{\mathscr{N}_0 m_e c^2}
\label{eqIC4}
\ ,
\end{equation}
where $\bar x_0$ denotes the corresponding initial value for the mean dimensionless momentum.\\

Solving Equation~(\ref{eq8}) subject to the initial condition given by Equation~(\ref{eqIC1}) yields the exact solution for the electron distribution during the rising phase of the transient. The result obtained is
\begin{equation}
N_{\rm rise}(x,y) = \frac{J_0}{\sigma \sqrt{2\pi}}
\left[\frac{x_0(x,y)}{x}\right]^{\hat C}
\left[\frac{1-\hat S x_0^2(x,y)}{1-\hat S x^2}\right]^{1-\frac{\hat C}{2}}
{\rm exp}\left\{-\frac{[\mu + x_0(x,y)]^2}{2\sigma^2}\right\} \ , \ \ x_{\rm min}(y)< x<\frac{1}{\sqrt{\hat{S}}}
\ ,
\label{eq11}
\end{equation}
where the minimum energy at dimensionless time $y$ is given by
\begin{equation}
x_{\rm min}(y) = \frac{1}{\sqrt{\hat S}} \, {\rm tanh}\left(y\sqrt{\hat S}\right)
\label{eq13}
\ ,
\end{equation}
and the injection energy for particles with current energy $x$ at time $y$ is computed using
\begin{equation}
x_0(x,y) = \frac{1}{\sqrt{\hat S}} \, {\rm tanh}\left[{\rm tanh}^{-1}(x \sqrt{\hat S})-y\sqrt{\hat S}\right]
\label{eq12}
\ .
\end{equation}
The parameter $\hat{C}$ appearing in Equation~(\ref{eq11}) is related to the shock-regulated escape process (see Equation~(32) from Paper~1). The exact solution for the electron distribution during the rising phase of the event given by Equation~(\ref{eq11}) can be used to compute a series of time-dependent theoretical $\gamma$-ray synchrotron spectra for comparison with a sequence of observational spectral data for a given sub-flare or wave, up to the peak time, $t=t_{\rm pk}$, or equivalently, $y=y_{\rm pk}$. The synchrotron spectra thus obtained are self-consistent since synchrotron energy losses are included in the electron transport equation. In order to complete the theoretical picture, we must also employ the solution for the electron distribution during the decaying phase of the event, as discussed below.

\subsection{Decaying Phase}
\label{secDecayingPhase}

The determination of the electron distribution during the decaying phase of the event requires us to solve Equation~(\ref{eq8}), combined with an appropriate initial condition, while also setting the escape time $t_{\rm esc} = t_{\rm ad}$ using Equation~(\ref{tesc2}). The initial condition is given by the rising phase solution (Equation~(\ref{eq11})) evaluated at the peak of the sub-flare or wave; that is,
\begin{equation}
N_{\rm decay}(x,y_{\rm pk}) = N_{\rm rise}(x,y_{\rm pk}) \ ,
\label{eq14}
\end{equation} 
where $y=y_{\rm pk}$ at the peak of the event. Equation~(\ref{eq14}) can be combined with Equation~(\ref{eq8}) to obtain the exact solution
\begin{multline}
N_{\rm decay}(x,y) = \frac{J_0 e^{-(t-t_{\rm pk})/t_{\rm ad}}}
{\sigma \sqrt{2\pi}}
\left[\frac{1-\hat S x_0^2(x,y)}{1-\hat S x^2}\right]
{\rm exp}\left\{-\frac{\left[\mu+x_0(x,y)\right]^{2}}{2\sigma^2}\right\}
\\
\times \left[\frac{1-\hat S x_0^2(x,y-y_{\rm pk})}{1-\hat S x_0^2(x,y)}\right]^{\hat C/2}
\left[\frac{x_0(x,y-y_{\rm pk})}{x_0(x,y)}\right]^{-\hat C}\ , \qquad x_{\rm min}(y)< x<\frac{1}{\sqrt{\hat{S}}} \ ,
\ \ y \ge y_{\rm pk} \ ,
\label{eq15}
\end{multline}
where $x_0(x,y)$ and $x_0(x,y-y_{\rm pk})$ are computed using Equation~(\ref{eq12}). Taken together, Equations~(\ref{eq11}) and (\ref{eq15}) provide the exact solution for the electron number distribution during the rising and decaying phases of a single sub-flare or wave. The global solution is therefore given by
\begin{equation}
N(x,y) = \begin{cases}
N_{\rm rise}(x,y) \ , &y_{\rm pk} \ge y \ge 0\ , \cr
N_{\rm decay}(x,y) \ , &y \ge y_{\rm pk} \ . \cr
\end{cases}
\label{eqGlobal}
\end{equation}

\subsection{Synchrotron Emission}

The synchrotron spectra and light curves computed using the electron distribution represented by Equation~(\ref{eqGlobal}) is self-consistent since synchrotron losses are included in the transport equation that we have solved. We assume here that the electron distribution is isotropic in the frame of the bulk flow, so that the synchrotron emission can  be computed by convolving the electron number distribution function (Equation~(\ref{eqGlobal})) with the single-particle synchrotron emission function \citep[e.g.,][]{becker92,kroon16}
\begin{equation}
Q_\nu(\nu,\gamma) = \frac{\sqrt{3}\,q^3 B}{m_e c^2}R
\left(\frac{\nu}{\gamma^2 \nu_s}\right) \ \ \propto \ \ {\rm erg \ s^{-1}
\ Hz^{-1}}
\ ,
\label{eq40old}
\end{equation}
where
\begin{equation}
\nu_s \equiv \frac{3 q B}{4 \pi m_e c}
\ ,
\label{eq41old}
\end{equation}
and \citep{crusius86}
\begin{equation}
R(x) =
\frac{x^2}{2}K_{4/3}\Big(\frac{x}{2}\Big)K_{1/3}\Big(\frac{x}{2}\Big)-
\frac{3x^3}{20}\Big[K^2_{4/3}\Big(\frac{x}{2}\Big)-
K^2_{1/3}\Big(\frac{x}{2}\Big)\Big]
\ .
\label{eq42old}
\end{equation}
The functions $K_{4/3}(x)$ and $K_{1/3}(x)$ represent modified Bessel functions of the second kind.
The corresponding total synchrotron power per unit frequency emitted by the isotropic electron distribution in the frame of the plasma is obtained by integrating over the electron number distribution, yielding
\begin{equation}
P_\nu(\nu,t) = \int_{x_{\rm min}[y(t)]}^\infty N(x,t) \,
Q_\nu(\nu,x) \, dx \ \propto \ {\rm erg \ s^{-1} \ Hz^{-1}}
\ ,
\label{eq16}
\end{equation}
with the corresponding observed flux density
\begin{equation}
\mathscr{F}_\nu(\nu,t) = \frac{P_\nu(\nu,t)}{4 \pi D^2}
\ \propto \ {\rm erg \ s^{-1} \ cm^{-2}\ Hz^{-1} }\ ,
\label{eq44old}
\end{equation}
where $D$ is the distance to the source and $P_\nu(\nu,t)$ is evaluated using Equation~(\ref{eq16}). The theoretical light curve can be computed by integrating the flux with respect to the photon frequency $\nu$.

\section{Application to 2007 September Event}
\label{application}

In this Section, we apply our model to the 2007 September $\gamma$-ray flare observed by {\em AGILE} and refer to \citet{striani13} for the various data sets. Following \citet{striani13}, we will use the terms ``wave'' and ``sub-flare'' to describe the two different types of temporal events observed during the 2007 September $\gamma$-ray flare. Waves are defined as events lasting about one week, whereas sub-flares last about a day or less. Hence these two types of events are qualitatively different, and therefore it is important to test our theoretical model by determining if it is capable of accounting for the spectral data and the light curves for both types of transients. \citet{striani13} provide spectra and light curves for the W1 wave and the sub-flares F1, F2, and F3. In this paper, we will focus on the W1 and F2 events since these have the highest quality of the available spectral data.\\

By utilizing our time-dependent model to analyze the spectral data for the 2007 September flare, we can extract a number of interesting physical quantities and functions such as the variable electric and magnetic field strengths, the variable electron energy distribution, the initial blob energy, the total radiated synchrotron energy, and the total event energy. We apply our model to the interpretation of the data for the W1 wave and the F2 sub-flare events in Sections~(\ref{subflare}) and (\ref{wave}), respectively.

\subsection{F2 Sub-Flare}
\label{subflare}

The light curve plotted in Figure~1 from \citet{striani13} depicts 12-hour binning of the flaring activity from MJD 54360-54390 (2007 September 27 - October 17), which comprises two waves and three sub-flares. The waves are plotted in black and labeled either W1 or W2. These waves have durations of several days, and hence they are clearly resolved using 12 hour time bins. On the other hand, the sub-flares F1, F2, and F3 are very short-duration episodes of enhanced emission, which are barely resolved by the 12 hour time bins. Hence no time-dependent spectra are available for the sub-flares, but instead only the peak spectra are available and therefore we will compare our model predictions with the peak spectra. The brightest sub-flare is the F2 event, and the peak spectrum for this feature is plotted in Figure~5 from \citet{striani13}. We can use our model to interpret the F2 sub-flare by computing the peak spectrum using Equation~(\ref{eq16}) and then comparing the result with the corresponding {\em AGILE} spectrum taken from \citet{striani13}. The theoretical parameters are varied until an acceptable qualitative fit is obtained.\\

\begin{figure}[h!]
\vspace{0.0cm}
\centering
\includegraphics[height=6cm]{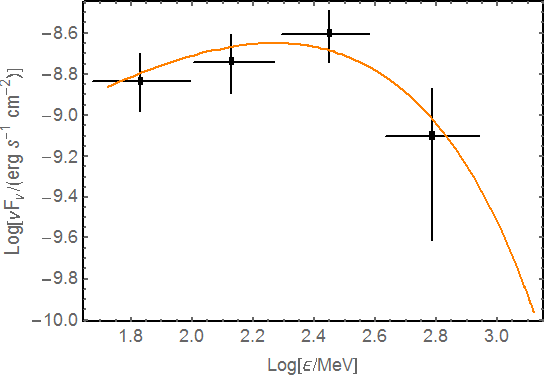}
\caption{The instantaneous peak spectrum of the F2 sub-flare, computed using Equation~(\ref{eq44old}) and evaluated at MJD 54382.40 (2007 October 9). The spectral data are from Figure~5 of \citet{striani13}.}
\label{fig1}
\end{figure}

\begin{figure}[h!]
\vspace{0.0cm}
\centering
\includegraphics[height=6cm]{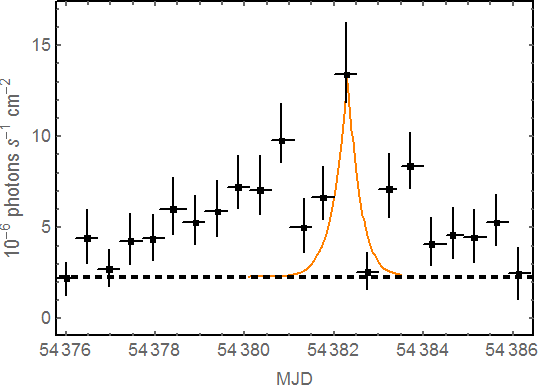}
\caption{The integrated light curve of the F2 sub-flare, for photon energies $> 100\,$MeV, computed using Equation~(57) from Paper~1, and compared with data taken from Figure~1 of \citet{striani13}.}
\label{fig2}
\end{figure}

In Figure~\ref{fig1}, we plot the theoretical peak spectrum, $\mathscr{F}_{\nu}(\nu,t_{\rm pk})$, for the F2 event computed using Equation~(\ref{eq44old}) along with the observational data taken from Figure~5 in \citet{striani13}, based on the theoretical parameter values listed in Tables~\ref{tbl-1} and \ref{tbl-2}. The peak magnetic field for the F2 sub-flare is $B_{\rm pk} = 713 \, \mu$G, and the event exhibits strong electrostatic acceleration, with a peak electric to magnetic field ratio $E_{\rm pk}/B_{\rm pk} = 1.23$. The peak electric field strength is $E_{\rm pk} = 877 \, \mu$G. The 12-hour duration of this sub-flare and the significant induced electrostatic fields at the peak imply that a substantial explosive reconnection event is taking place at the termination shock. The associated light curve above 100\,MeV is compared with the observational data in Figure~\ref{fig2}. It is important to note that according to Figure~1 from \citet{striani13}, the rising side of the F2 sub-flare is ``contaminated'' by the decaying side of the F1 sub-flare. Since the F1 feature is not modeled here (due to a lack of spectral data), we only plot the theoretical light curve for F2. Hence the theoretical light curve slightly underestimates the observed flux on the rising side of F2, as expected.

\subsection{W1 Wave}
\label{wave}

During the 2007 September flare from the Crab Nebula, the W1 wave event persisted over several days, during the time interval MJD 54368-54373 (2007 September 25-30). \citet{striani13} provide a single integrated spectrum for the W1 wave in their Figure~5, but a true time-dependent spectrum for this feature is not currently available in the literature. The corresponding light curve for the W1 event is depicted in both Figures~1 and 4 from \citet{striani13} using 12-hour and 24-hour time binning, respectively. The peak is clearly resolved in their Figure~1 but not in Figure~4 due to the coarse time bins. In this study, we will therefore test our model using the 12-hour binned light curve for W1 taken from Figure~1 of \citet{striani13}.\\

\begin{figure}[h!]
\vspace{0.0cm}
\centering
\includegraphics[height=6cm]{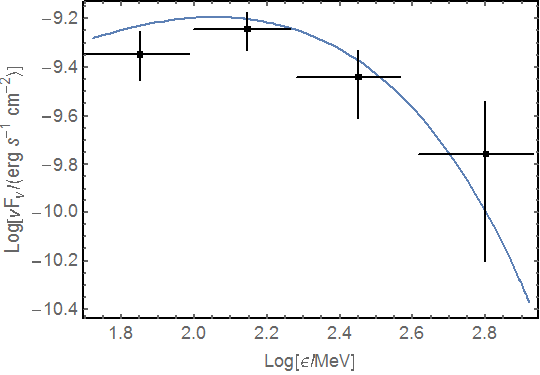}
\caption{The time-averaged spectrum of the W1 wave transient, computed using Equation~(\ref{eq17}) and integrated from MJD 54368-54373 (2007 September 25-30). The spectral data are from Figure~5 of \citet{striani13}.}
\label{fig3}
\end{figure}

\begin{figure}[h!]
\vspace{0.0cm}
\centering
\includegraphics[height=6cm]{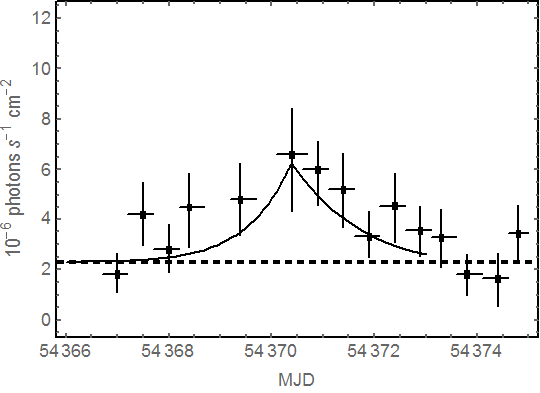}
\caption{The integrated light curve of the W1 wave transient, for photon energies $> 100\,$MeV, computed using Equation~(57) from Paper~1, and compared with data taken from Figure~1 of \citet{striani13}.}
\label{fig4}
\end{figure}

Although our theoretical model is fully time-dependent, in the case of the W1 transient, we must perform a time average over the interval MJD 54368-54373 in order to make contact with the spectrum for this event plotted in Figure~5 from \citet{striani13}. Hence the theoretical spectrum for the W1 event is obtained using the time integration
\begin{equation}
\bar{\mathscr{F}}_{\nu}(\nu) = \dfrac{1}{t_2-t_1}\int_{t_1}^{t_2}\mathscr{F}_{\nu}(\nu,t) \, dt
\ ,
\label{eq17}
\end{equation}
where $t_1$ and $t_2$ are the starting and ending times of the W1 flare, and $\mathscr{F}_{\nu}(\nu,t)$ is computed using Equation~(\ref{eq44old}). The resulting spectrum, $\bar{\mathscr{F}}_{\nu}(\nu)$, is compared with the observed time-averaged spectrum for W1 taken from Figure~5 in \citet{striani13}, and the theoretical parameters are varied until adequate agreement is obtained. Our results for the theoretical W1 spectrum are plotted along with the observational data in Figure~\ref{fig3}, and the corresponding light curve above 100\,MeV is plotted and compared with the observational data in Figure~\ref{fig4}. The W1 wave event is not as bright at its peak as the F2 sub-flare, and therefore the peak magnetic field for W1 is only $B_{\rm pk} = 472 \, \mu$G. Further parameters are provided in Tables~\ref{tbl-1} and \ref{tbl-2}.

\section{Discussion and Conclusion}
\label{conclusion}

The remarkable series of bright $\gamma$-ray flares emitted by the Crab Nebula and observed by {\em AGILE} or {\em Fermi} between 2007-2013 has presented considerable challenges to classical particle acceleration models, due in part to the apparent violation of the classical synchrotron burn-off limit, implying particle acceleration on sub-Larmor timescales. We have previously demonstrated in Paper~1 that the time-dependent spectra observed during the intense 2011 April $\gamma$-ray flare can be described using a relatively simple analytical model. A time-independent version of the model was also compared with all of the available flare data by \citet{kroon16}. While the earlier model comparisons have been promising, the 2007 September Crab Nebula $\gamma$-ray flare was a unique event since it comprised both short-duration sub-flares of the sort that have been seen during other transients, combined with long-duration wave events lasting about one week. Hence the analysis of these data represents the strongest test of the theoretical model we have developed. This has motivated us to test our model by making detailed comparisons with all of the available data for the 2007 September event, as presented by \citet{striani13}. We discuss our primary findings below.\\

The 2011 April light curve is characterized by two distinct peaks, known as sub-flares \citep{buehler12}, which in our model are treated as emission from two individual blobs that pass separately through the termination shock, but are both included in the same overall flare event. The first 2011 sub-flare was found to have a peak magnetic field $B_{\rm pk} = 706 \, \mu$G and an electric/magnetic field ratio $E_{\rm pk}/B_{\rm pk}=1.84$ at the peak. These values are similar to those obtained here for the F2 sub-flare observed during the 2007 event, for which we obtain a peak magnetic field $B_{\rm pk} = 713 \, \mu$G and a peak field ratio $E_{\rm pk}/B_{\rm pk} = 1.23$. In our analysis of the 2007 September flare, we find that the mean Lorentz factor for the incident electrons is $\bar\gamma_0 \sim 10^9$ (see Table~2), which is significantly larger than the upstream Lorentz factor in the cold pulsar wind, $\Gamma \sim 10^6$ \citep{cerutti14}. This implies that the electrons in the incident plasma blob are produced as a result of impulsive reconnection on the upstream side of the termination shock \citep{montani14}. The values for the magnetization parameter, $\sigma$, obtained here are $\sigma=0.036$ for W1 and $\sigma=7.4 \times 10^{-5}$ for F2, which agree with the expected values in the downstream region \citep{sironi09}. On the other hand, PIC simulations carried out by \citet{lyutikov17,lyutikov18} and \citet{cerutti14} suggest that $\sigma \gg 1$. This is known as the ``$\sigma$-problem'' in pulsar astrophysics and is currently unresolved \citep{buehler14,lyutikov18}.\\

\citet{striani13} posit that sub-flares are powered by emission from blobs with radii $R_{\rm b} \sim (1-5) \times 10^{15}\,$cm possessing relatively strong magnetic fields, on the order of $10^3\,\mu$G. Conversely, the waves are associated with larger plasma blobs with radii $R_{\rm b} \gapprox 10^{16}\,$cm, possessing weaker magnetic fields that the sub-flare blobs. It is interesting to compare the estimates obtained by \citet{striani13} with the physical parameters resulting from the application of our model to the same event. During the sub-flare F2, we find a blob radius $R_{\rm b} = 10^{15}$ cm and a peak magnetic field $B_{\rm pk} = 713 \, \mu$G. By contrast, during the W1 wave we obtain a blob radius $R_{\rm b} = 10^{16}$ cm and a peak magnetic field $B_{\rm pk} = 472 \, \mu$G. These values agree reasonably well with the estimates obtained by \citet{striani13}, which are $(1.5 \pm 0.5) \times 10^3 \, \mu$G for F2 and $(0.8 \pm 0.2) \times 10^3 \, \mu$G for W1. However, it is important to emphasize that the parameters obtained by \citet{striani13} were based on an analysis of the energy budget for the event, whereas our parameters are obtained via detail spectral comparisons. Hence the two determinations are separate but complementary, and the fact that they agree strongly supports the interpretation that the sub-flares and waves are powered by explosive reconnection.\\

\begin{figure}[h!]
\vspace{0.0cm}
\centering
\includegraphics[height=6cm]{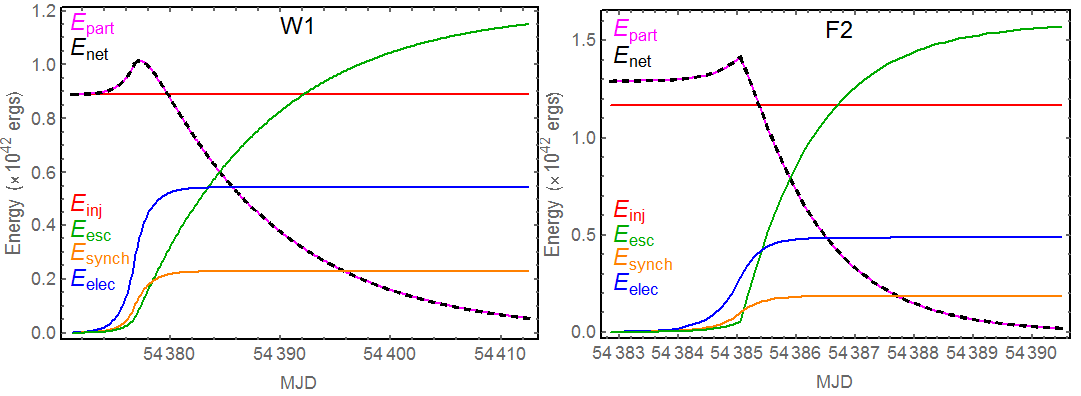}
\caption{The time-dependent cumulative energy produced in each channel, computed using Equations~(58)-(65) from Paper~1.}
\label{fig5}
\end{figure}

 The ``sub-flare/wave'' paradigm adopted by \citet{striani13} represents an interesting classification scheme, since the waves tend to be longer-lasting but less intense episodes of enhanced $\gamma$-ray activity, compared to the more powerful but brief ``sub-flare'' counterparts. This classification can be further explored by examining the energetics of the various temporal features observed during the 2007 flare. The energy integrals for each channel are defined in Equations~(58)-(65) from Paper~1, and are denoted by $\mathscr{E}_{\rm inj}$ (initial electron energy in blob), $\mathscr{E}_{\rm part}(t)$ (total electron energy in blob), $\mathscr{E}_{\rm elec}(t)$ (total electrostatic acceleration energy), $\mathscr{E}_{\rm synch}(t)$ (total energy emitted in synchrotron radiation), and $\mathscr{E}_{\rm esc}(t)$ (total energy in escaped electrons). We have computed the time-dependent energy in each of these channels and plotted them as functions of time for W1 and F2 in Figure~\ref{fig5}. The asymptotic values for each channel energy obtained as $t \to \infty$ are listed in Table~\ref{tbl-3}.\\

Despite the difference in duration between W1 and F2, we find that they each produce a total energy of $\sim 2 \times 10^{41}\,$erg in the synchrotron channel. The average power of the sub-flare F2 is about five times greater than that of the W1 wave due to its shorter duration. Figure~\ref{fig5} indicates that most of the energy advects downstream from the shock and goes on to power the synchrotron emission in the outer region of the nebula. Depending on the strength of the magnetic field in the nebula, this could result in an observable X-ray afterglow. In fact, \citet{kroon16} considered the afterglow associated with this theoretical model for each of the various Crab Nebula $\gamma$-ray flares. They found that an observable X-ray afterglow could persist for up to two weeks, depending on the magnetic field strength. The theoretical prediction for the 2007 September flare was plotted in Figure 5 from \citet{kroon16}, but no X-ray observations were performed after this particular $\gamma$-ray transient, and therefore the prediction cannot be compared with data.\\

The standard observational estimate for the efficiency of the conversion of the pulsar's spin-down power into synchrotron radiation is $\sim 30\%$ \citep{abdo11}. It is interesting to compare this estimate with the theoretical efficiency, defined as the ratio $\mathscr{E}_{\rm synch}/(\mathscr{E}_{\rm synch}+\mathscr{E}_{\rm esc})$, which can be evaluated using the asymptotic values for the channel energies listed in Table~\ref{tbl-3}. The resulting theoretical efficiencies obtained for the F2 and W1 events are $\sim 10\%$ and $\sim 15\%$, respectively, which are similar to the results obtained in our analysis of the 2011 April flare in Paper~1. The discrepancy between our results and the spin-down estimates could be due to the fact that the spin-down estimate is a time average, and therefore it does not accurately represent the energetics during an individual transient. The asymptotic values of the electrostatic energy channels in the W1 and F2 events are both $\sim 5 \times 10^{41}\,$erg, which is expected given the similar levels in the synchrotron energy channel.\\

It is interesting to compare the energy budget for the 2007 September event treated here with that from our previous analysis of the 2011 April flare in Paper~1. The total synchrotron energies in sub-flares 1 and 2 for the 2011 April event were found to be $4 \times 10^{41}$ and $7 \times 10^{41}$ erg, respectively, which exceed the synchrotron values derived here for W1 and F2 by a factor of $\sim 2$. Comparison of the energy in the electrostatic channel for each event can reveal useful information about the magnitude of the induced electrostatic fields primarily responsible for accelerating the electrons. For the 2007 September event, the maximum energy generated by the fields is seen to be about half of the initial energy in the blob for both W1 and F2. However, in the two sub-flares observed during the 2011 April event, we found in Paper~1 that the electrostatic energy considerably exceeds the initial energy of the blob, which implies a very energetic and efficient conversion of wind energy into synchrotron $\gamma$-rays. Hence the 2007 flare was a much more mild event. Another useful metric for establishing the importance of electrostatic acceleration is provided by the peak-field ratio $E_{\rm pk}/B_{\rm pk}$, which traces the explosiveness of magnetic reconnection during the transient. In the case of the W1 and F2 features, this ratio is of order unity (Table \ref{tbl-2}). However, during the more energetic and efficient sub-flares observed in 2011 April, we found that $E_{\rm pk}/B_{\rm pk} \sim 3$.\\

For the 2011 April sub-flares, our model indicates that the peak magnetic fields were in the range $B_{\rm pk} \sim 600-700 \, \mu$G (see Paper~1). These results are similar to those obtained here for the 2007 flare, which implies that the same particle acceleration and emission mechanisms are operating throughout the range of $\gamma$-ray flares observed from the Crab Nebula. We conclude that our model is able to reproduce the observational data ($\gamma$-ray spectra and integrated light curves) for both the 2007 September and the 2011 April $\gamma$-ray flares from the Crab Nebula, despite the fact that the energy budgets were distinctly different for these two events. Analysis of the 2007 September $\gamma$-ray flare is especially important because this event comprised qualitatively different temporal features compared with any of the other Crab Nebula flares. The ability of our model to account for both the long-duration waves and the short sub-flares suggests that our theoretical framework for treating particle acceleration in the explosive reconnection environment surrounding the termination shock can help to provide a detailed understanding of the physics taking place in these extreme astrophysical sources.

\acknowledgements
The authors are grateful to the anonymous referee for providing several useful comments that helped to strengthen and improve the presentation. J.J.K. was supported at NRL by NASA under contract S-15633Y. J.D.F. was supported by the Chief of Naval Research.\\

\hoffset=-0.88truein
\begin{deluxetable}{ccccccccccc}
\tabletypesize{\scriptsize}
\tablecaption{Model Free Parameters \label{tbl-1}}
\tablewidth{0pt}
\tablehead{
\colhead{\rm event}
& \colhead{$J_0$}
& \colhead{$\dfrac{E_*}{B_*}$}
& \colhead{$\hat S$}
& \colhead{$\hat C$}
& \colhead{$\mu$}
& \colhead{$\sigma$}
& \colhead{$\alpha$}
& \colhead{$\theta$}
& \colhead{$t_{\rm adv}\ (s)$}
& \colhead{$t_* \ ({\rm MJD})$}
}
\startdata
$W1$
&$1.86 \times 10^{39}$
&$0.06$
&$3.16 \times 10^{-20}$
&$0.2$
&$10^{5}$
&$1.00 \times 10^{10}$
&$5.81$
&$4.80$
&$1.01 \times 10^6$
&$54364.0$
\\
$F2$
&$3.24 \times 10^{39}$
&$0.04$
&$4.27 \times 10^{-20}$
&$0.2$
&$10^{5}$
&$8.91 \times 10^{9}$
&$6.85$
&$6.90$
&$1.08\times 10^{5}$
&$54379.8$
\\
\enddata
\end{deluxetable}

\hoffset=-0.88truein
\begin{deluxetable}{ccccccccccc}
\tabletypesize{\scriptsize}
\tablecaption{Derived Parameters \label{tbl-2}}
\tablewidth{0pt}
\tablehead{
\colhead{\rm event}
& \colhead{$\mathscr{N}_0$}
& \colhead{$\bar{\gamma}_0$}
& \colhead{$R_{\rm b} \ ({\rm cm})$}
& \colhead{$B_{\rm pk} \ (\mu{\rm G})$}
& \colhead{$\dfrac{E_{\rm pk}}{B_{\rm pk}}$}
& \colhead{$A_* \ ({\rm s}^{-1})$}
& \colhead{$w_{*}$}
& \colhead{$w_{\rm pk}$}
& \colhead{$t_{\rm pk} \ ({\rm s})$}
}
\startdata
$W1$
&$3.97 \times 10^{38}$
&$2.36 \times 10^{9}$
&$1.01 \times 10^{16}$
&$472$
&$1.10$
&$27.3$
&$83.3$
&$4.56$
&$5.52 \times 10^{5}$
\\
$F2$
&$6.68 \times 10^{38}$
&$2.74 \times 10^{9}$
&$1.08 \times 10^{15}$
&$713$
&$1.23$
&$16.3$
&$125$
&$4.07$
&$2.38 \times 10^{5}$
\\
\enddata
\end{deluxetable}

\hoffset=-0.88truein
\begin{deluxetable}{cccccc}
\tabletypesize{\scriptsize}
\tablecaption{Asymptotic Energy Values \label{tbl-3}}
\tablewidth{0pt}
\tablehead{
\colhead{\rm event}
& \colhead{$\mathscr{E}_{\rm inj}$ (erg)}
& \colhead{$\mathscr{E}_{\rm elec}$ (erg)}
& \colhead{$\mathscr{E}_{\rm synch}$ (erg)}
& \colhead{$\mathscr{E}_{\rm esc}$ (erg)}
}
\startdata
$W1$
&$8.89 \times 10^{41}$
&$5.65 \times 10^{41}$
&$2.40 \times 10^{41}$
&$1.10 \times 10^{42}$
\\
$F2$
&$1.17 \times 10^{42}$
&$5.45 \times 10^{41}$
&$2.12 \times 10^{41}$
&$1.52 \times 10^{42}$
\\
\enddata
\end{deluxetable}

\end{document}